\documentclass[prd,onecolumn,nofootinbib]{revtex4}
\usepackage{bm,amsmath,amssymb,graphicx}

\begin{document}

\title{Invariant temperature of a moving body}
\author{Nikodem Pop{\l}awski}
\affiliation{Department of Mathematics and Physics, University of New Haven, 300 Boston Post Road, West Haven, CT 06516, USA}
\email{NPoplawski@newhaven.edu}

\begin{abstract}
The temperature of a mechanical body has a kinetic interpretation: it describes the relative motion of particles within the body.
Since the relative velocity of two particles is a Lorentz invariant, so is the temperature. 
In statistical physics, the temperature is defined as the inverse of the partial derivative of the entropy with respect to the internal energy (the energy in the rest frame of reference).
Since the internal energy is a Lorentz invariant, so is the temperature.
The Lorentz invariance of the temperature is a consequence of the symmetry between two bodies with equal proper temperatures, moving relative to one another with a constant relative speed, and in thermal contact.
We give an equivalent, covariant definition of the temperature in terms of the energy and momentum of the body.
We also note contradictions in the earlier articles that derived various transformation laws for the internal energy and temperature.
\end{abstract}
\maketitle

The temperature of a body moving with a constant velocity has been the subject of a long and interesting debate, as reported in a recent review \cite{rev}.
Planck \cite{Pla} and Einstein \cite{Ein} derived that a body, whose proper temperature (in its rest frame of reference) is $T_0$, has the temperature
\begin{equation}
T=T_0/\gamma
\label{colder}
\end{equation}
in a frame of reference in which it is moving with a speed $v$, with $\gamma=(1-v^2/c^2)^{-1/2}$.
Accordingly, a moving body would appear colder.
This formula has been adopted as the orthodox transformation \cite{cold}.
Ott \cite{Ott} derived another formula for the temperature of a moving body:
\begin{equation}
T=T_0\gamma.
\label{hotter}
\end{equation}
Accordingly, a moving body would appear hotter \cite{Arz,hot,Sut,New}.
Some authors accepted both formulae depending on additional criteria \cite{two}.
Landsberg \cite{Lan1,Lan2} proposed that the temperature should be an invariant under Lorentz transformations \cite{invar}:
\begin{equation}
T=T_0.
\label{same}
\end{equation}
Cavalleri and Salgarelli \cite{CS} suggested that temperature is meaningful only in the rest frame and thus the Lorentz transformation for temperature and other thermodynamic quantities does not exist \cite{equal}.

The formulae (\ref{colder}), (\ref{hotter}), and (\ref{same}) give different transformation laws for the temperature \cite{Kam}, although they all were derived using the invariance of the entropy.
Landsberg pointed out \cite{Lan1} that the original formula (\ref{colder}) used the following definition of temperature:
\begin{equation}
\frac{1}{T}=\Bigl(\frac{\partial S}{\partial E}\Bigr)\Big|_V,
\label{ener}
\end{equation}
where $V$ is the volume of the body, $E$ is its energy, and $S$ is its entropy.
However, in the kinetic interpretation, the temperature of a body should be a measure of the average kinetic energy of the relative motion of particles in the body.
Since the relative velocity of two particles is an invariant, the temperature should be invariant as well.
He proposed that the temperature should be defined in statistical physics as \cite{Lan1}
\begin{equation}
\frac{1}{T}=\Bigl(\frac{\partial S}{\partial U}\Bigr)\Big|_V,
\label{temp}
\end{equation}
where $U$ is its internal energy.
Since the internal energy of a body is defined as the energy of the body in its rest frame of reference, in which the momentum of the body is zero \cite{LL1}, it is invariant by construction.
It is given by \cite{Lan1}, 
\begin{equation}
U=[E^2-(c{\bf P})^2]^{1/2},
\label{inten}
\end{equation}
where ${\bf P}$ is the momentum of the body.
The internal energy is thus equal to the invariant mass of the system \cite{LL2}.
Also, the entropy as the logarithm of the number of microstates corresponding to the body macrostate is invariant.
Consequently, the temperature defined in equation (\ref{temp}) is invariant and equal to $T_0$.
The resulting formula (\ref{same}) represents the correct transformation law for the kinetic temperature.
The formulae (\ref{colder}) and (\ref{hotter}) do not represent such a transformation law.

A simple and natural justification of the Lorentz invariance of the temperature, supporting the formula (\ref{same}), was given by Landsberg \cite{Lan2}.
He considered two bodies in thermal equilibrium, with equal proper temperatures $T_0$, moving relative to one another with a constant relative speed $v$, that have parts of their surfaces near each other for some period of time.
Assuming the orthodox transformation, the temperature of a moving body is given by (\ref{colder}).
In the frame of reference of the first body, the second one is moving with speed $v$.
In this frame, $T_1=T_0$ and $T_2=T$, thereby the second body is colder than the first one.
According to the second law of thermodynamics, heat can flow spontaneously only from a hotter body to a colder body.
Therefore, heat flows from the first body to the second one.
Consequently, the temperature of the first body is decreasing and the temperature of the second one is increasing.

In the frame of reference of the second body, the situation is opposite: the first body is moving and its temperature $T_1=T$ is lower than the temperature of the second body $T_2=T_0$.
Accordingly, heat flows from the second body to the first one, the temperature of the first body is increasing and the temperature of the second one is decreasing.
However, the temperature of a body cannot increase in one frame and decrease in another (otherwise, a phase transition could happen in one frame but not in another, violating the physical equivalence of inertial frames).
Also, the sign of heat flow between two systems is invariant \cite{Red} (the amount of heat flow depends on the frame of reference \cite{Wil}).
Therefore, the assumption (\ref{colder}) leads to a contradiction.
A similar contradiction is reached by assuming the alternative transformation (\ref{hotter}) or, in general, any transformation law $T=T_0 f(v)$ with an arbitrary function $f$, unless $f=1$.
The presented argument leads to the conclusion that $T=T_0$ \cite{Lan2}.
The temperature of a moving body is equal to its temperature in the rest frame of reference, and thus is a Lorentz invariant.

We propose a covariant definition of the temperature of a body with energy $E$ and momentum ${\bf P}$, moving with velocity ${\bf v}={\bf P}c^2/E$:
\begin{equation}
\frac{1}{T}=\Bigl[\Bigl(\frac{\partial S}{\partial E}\Bigr)\Big|_V^2-\Bigl(\frac{\partial S}{c\,\partial {\bf P}}\Bigr)\Big|_V^2\Bigr]^{1/2}.
\label{covtemp}
\end{equation}
This quantity can also be written as
\begin{equation}
\frac{1}{T}=\Bigl[g_{ij}\Bigl(\frac{\partial S}{c\,\partial P_i}\Bigr)\Big|_V\Bigl(\frac{\partial S}{c\,\partial P_j}\Bigr)\Big|_V\Bigr]^{1/2},
\label{temptens}
\end{equation}
where $g_{ij}$ is the Minkowski metric tensor and $P_i$ is the covariant four-vector of momentum formed by the time component $P_0=E/c$ and the space components $-{\bf P}$ \cite{LL2}.
The components $(\partial S/c\partial P_i)|_V$ form a contravariant four-vector, whose four-dimensional magnitude is equal to $1/T$.
The temperature (\ref{temptens}) is clearly invariant.
It is also equivalent to (\ref{temp}):
\[
\frac{1}{T}=\Bigl(\frac{\partial S}{\partial U}\Bigr)\Big|_V\Bigl[\Bigl(\frac{\partial U}{\partial E}\Bigr)^2-\Bigl(\frac{\partial U}{c\,\partial {\bf P}}\Bigr)^2\Bigr]^{1/2}=\Bigl(\frac{\partial S}{\partial U}\Bigr)\Big|_V\Bigl[\frac{E^2}{U^2}-\frac{c^2{\bf P}^2}{U^2}\Bigr]^{1/2}=\Bigl(\frac{\partial S}{\partial U}\Bigr)\Big|_V,
\]
where we used (\ref{inten}).
Consequently, this definition is consistent with the formula (\ref{same}).

We also note that the inverse of this temperature can be regarded as the magnitude of the four-vector $(1/c)\partial S/\partial P_i$.
For a mechanical body, $P_i=-\partial\mathfrak{S}/\partial X^i$, where $\mathfrak{S}$ is the action as a function of the coordinates $X^i$ \cite{LL2}.
Accordingly, we have
\begin{equation}
\frac{1}{T}=\Bigl[g_{ij}\Bigl(\frac{\partial S}{c\,\partial(\partial\mathfrak{S}/\partial X^i)}\Bigr)\Big|_V\Bigl(\frac{\partial S}{c\,\partial(\partial\mathfrak{S}/\partial X^j)}\Bigr)\Big|_V\Bigr]^{1/2}.
\end{equation}
In this definition, $X^i$ are the coordinates of an extended body in the prescription of Papapetrou \cite{Pap}.

The internal energy $U$, equal to the energy in the rest frame of reference, is invariant by construction: $U=U_0$.
Consequently, the heat $Q=\Delta U|_V$ is also invariant: $Q=Q_0$.
The thermodynamic pressure is defined as $p=-(\partial U/\partial V_0)|_S$, and is invariant by construction: $p=p_0$.

We note that the transformation laws for the internal energy and heat given in the earlier articles, leading to the frame dependence of the temperature, are incorrect.
The orthodox transformation (\ref{colder}) was derived from transformation laws $U=\gamma(U_0+\beta^2 p_0 V_0)$, $U+pV=\gamma(U_0+p_0 V_0)$, and $Q=Q_0/\gamma$, where $\beta=v/c$ and $p=p_0$.
They contradict the invariance of $U$ and $Q$.
Ott \cite{Ott} used $Q=Q_0\gamma$, which contradicts the invariance of $Q$.
Arzelies \cite{Arz} used $U=U_0\gamma$ and $Q=Q_0\gamma$, which contradict the invariance of $U$ and $Q$.
Sutcliffe \cite{Sut} used $U=U_0\gamma+f(\beta)$, where $f$ is some function, which contradicts the invariance of $U$.
Newburgh \cite{New} considered a body emitting heat isotropically in its rest frame of reference and concluded that the invariance of the radiated power yields the formula (\ref{hotter}).
However, the invariance of power $P=dQ/dt$ contradicts the invariance of $Q$.
Finally, the authors of the recent review \cite{rev} used (\ref{ener}) instead of the kinetically correct (\ref{temp}), and $Q=Q_0\gamma$ contradicting the invariance of $Q$.
As a result, they obtained the formula (\ref{hotter}).
However, the kinetic (physical) temperature is invariant, as given by the formula (\ref{same}).

This work was funded by the University Research Scholar program at the University of New Haven.

\end{document}